\documentclass[%
 reprint,
 amsmath,amssymb,
 aps,
]{revtex4-2}

\usepackage{graphicx}% Include figure files
\usepackage{dcolumn}% Align table columns on decimal point
\usepackage{bm}% bold math
\usepackage[dvipsnames]{xcolor}
\usepackage{comment}
\begin{document}

\preprint{APS/123-QED}

\title{Continuous  simultaneous measurement of  position and  momentum of a particle
}% Force line breaks with \\

\author{Filip Gampel}
\email{gampel@ifpan.edu.pl}
\author{Mariusz Gajda}%
 
\affiliation{%
 Institute of Physics, Polish Academy of Sciences, Aleja Lotników 32/46, 02-668 Warsaw, Poland
}%

\date{\today}% It is always \today, today,

\begin{abstract}
%We formulate a description of a quantum particle continuously monitored  by detectors measuring simultaneously its position and momentum. We  account for the back-action of the meters and assume that upon detection the particle is found in one of the meters' states chosen as a discrete subset of coherent states.  A system (particle) interacting with an environment (detectors) is described by the Gorini-Kossakowski-Sudarshan-Lindblad equation. We solve the equation  using the Wave-Function Quantum Monte Carlo approach which allows for interpretation of a sequence of measurements as a single trajectory of the particle.  The different regimes of dynamics of the free particle are quantitatively discussed: from stroboscopic motion in the case of low interrogation frequency, to delayed dynamics reminiscent of the Zeno effect, if monitoring is frequent. The emergence of classical trajectories for semi-continuous spatial distribution of meters is shown. Statistical properties of the trajectories are discussed. Finally a comparison to other detection schemes in which the effect of measurement corresponds to filtering is presented. 

We formulate a model of a quantum particle continuously monitored by detectors measuring simultaneously its position and momentum.  We implement the postulate of wavefunction collapse by assuming that upon detection the particle is found in one of the meters' states chosen as a discrete subset of coherent states. The dynamics, as observed by the meters, is thus a random sequence of jumps between coherent states. We generate such trajectories using the Monte Carlo Wavefunction method. For sparsely distributed detectors, we use methods from renewal theory of stochastic processes to obtain some semi-analytic results. In particular, the different regimes of dynamics of the free particle are identified and quantitatively discussed: from stroboscopic motion in the case of low interrogation frequency, to delayed dynamics reminiscent of the Zeno effect if monitoring is frequent. For a semi-continuous spatial distribution of meters the emergence of classical trajectories is shown. Their statistical properties are discussed and compared to other detection schemes in which the effect of measurement corresponds to "spatial filtering" of the wavefunction.

\end{abstract}

\maketitle

\section{Introduction}
For more than a century the principles of quantum mechanics are employed successfully in the study of an immensely broad range of phenomena and it forms a basis of the modern physical worldview. The theory is developed from a set of axioms including, at least in the conventional Copenhagen formulation \cite{Born1927,Heisenberg1930}, the postulate of wavefunction collapse: upon observation, the pre-measurement state of the system immediately changes to the projection onto the eigensubspace associated with the measurement's outcome \cite{Fine69,vonNeumann32,Luders51, Khrennikov09}. From its outset this rule sparked an ongoing controversy, and as part of the \emph{measurement problem} it remains a subject of debate to the present day. Measurement grows into the central issue in most attempts to understand the emergence of the classical from the quantum \cite{Zurek2003}. It is often treated as a destructive process performed repeatedly on an ensemble of identically prepared quantum states. The problem of continuous monitoring of a single quantum system was considered merely of academic interest by the founding fathers of quantum mechanics since in these days it was inconceivable to experimentally study an isolated quantum system in a single (non-averaged) realization.\\

Single quantum systems suitable for controlled long-time observations appeared with the first trapped ions \cite{Toschek78}. This experimental achievement triggered theoretical advances in the field of repeated measurement of optical transitions in a single quantum system \cite{Wilkens97}. Advances in techniques of manipulating and observing atoms in periodic lattices, in particular the construction of atomic microscopes, allowed for direct observation of the motion of individual quantum objects in present-day experiments with ultracold atoms \cite{Bakr09,Bakr10,Sherson10,Cheuk15,Parsons15,Haller15,Edge15}. One can directly monitor atoms illuminated by light and jumping between adjacent sites in optical lattices. These experiments raised again some interest in the effect of continuous observation of quantum systems \cite{Javanainen13,Ueda17,Spielman19,Spielman21}.\\

One of the motivations for the present study stems from the work on single-shot measurements of many-body quantum systems \cite{Zaluska16,Gajda17} where high-order spatial correlations  may reveal distinct geometric structures. Similarly, simultaneous detection of many particles unveils the solitonic nature of type-II excitation in the Lieb-Linniger model on a ring \cite{Syrwid15,Syrwid16}. In both these cases the one-particle density does not indicate any of said structures. They emerge only in the course of multipoint measurement on the same system. A similar effect, but in the time domain, is responsible for the phenomenon of time crystals  as envisioned by Wilczek \cite{Wilczek12} (see also \cite{Sacha15,Zakrzewski18,Sacha20}). Breaking the time-translational symmetry of stationary states initiates a periodic motion of the system. Repeated measurement of {\it the  same} system is necessary to prove this periodic motion. If the sequence of measurements was performed each time on a newly prepared system  some chaotic rather than periodic motion would be observed. Every measurement on a newly prepared system breaks the symmetry differently, so the initial position of the system  will vary randomly from one realization to the other.\\ 

To account for the effects mentioned above a back-action of the meters must be incorporated into the description. Canonical theory of quantum measurement \cite{vonNeumann32,London39,Wigner63,Arthurs65,Caves87,Milburn01} assumes a quantum system coupled to a meter. The interaction between the two introduces correlations, the meter wavefunction becomes entangled with the system; then a projection operator corresponding to the observable of interest is applied to readout the meter. If the measurement is described in terms of mapping on a complete set of self-adjoint projectors then it constitutes a projection-valued measure (PVM).\\    
%prepared  typically in the state $\propto e^{-x^2/2\sigma}$. The coupling between the system and the meter is often assumed \cite{vonNeumann32,Arthurs65,Caves87,Milburn01} to be of the form $H_{int}=\gamma A \otimes {\hat p}$, where $\gamma$ is the interaction strength, $A$ is the observable to be measured and $\hat x$ and $\hat p$ are canonical position and momentum of the meter. Upon the unitary evolution, 
%$U=e^{-iH_{int}t/\hbar} = e^{-i \gamma A \otimes {\hat p}t/\hbar }$ the meter wavefunction becomes entangled with the system and is shifted by a value proportional to the measured observable. 

Historically, continuous measurements of quantum objects were first considered in 1969 by Davies  as a quantum stochastic process  in the context of photon counting \cite{Davies69}. Later, description of continuous measurement in the framework of path integrals formalism was suggested by Mensky \cite{Mensky79}. This approach was further developed by Caves \cite{Caves87}. Observables with continous spectra, like the position operator, were considered by  Barchielli, Lanz and Prosperi \cite{Barchielli82,Barchielli84} (see also \cite{Barchielli83,Barchielli85,Barchielli86,Barchielli12}) who  instead of projection operators   introduced the more general concepts of "effects" and "operations". "Effect" (positive operator valued measure) is a generalization of the notion of observable, typically viewed as a projective-valued measure. The term "operation" generalizes the reduction of the wavepacket \cite{ Barchielli83, Barchielli85, Barchielli86, Barchielli12, Kraus83}. Most approaches are based on the stochastic Schr\"odinger equation or on stochastic quantum differential equations for the density operator, \cite{Caves87, Milburn01, Ghirardi86, Gisin84, Diosi88a, Diosi88b} describing a system which experiences some external noise. \\

Investigations of measurement of observables with continous spectra lead to the formulation of a theory where the effect of measurement of a particle is accounted for by multiplying its wavefunction by a meter-dependent envelope - a filter localized at some spatial position \cite{Ueda17,Barchielli82,Ghirardi86,Caves87}. This kind of description departs from the orthodox postulate of the collapse of the wavefunction onto the meter state. The localization scheme does not affect significantly the wavefunction locally at the position of the meter $x_0$, but filters the wavefunction by diminishing distant amplitudes. In particular the filtering does not change substantially the local phase $\varphi(x_0)$ of the wavefunction i.e. the local velocity $v(x_0) \propto \partial_x \varphi(x_0)$ is "almost" not modified. The post-measurement state of the system is not completely determined by the meter, it preserves some information of its pre-measurement history. Nevertheless, in the words of Caves \cite{Caves87}, this filtering-type of localization {\it "nowhere invokes wave-function collapse, yet hidden within it must be a way of thinking in terms of collapse"}.  The projective measurement is performed on the system's environment. \\

In this paper we want to present an approach rooted in standard quantum mechanics which is capable of tracing the dynamics of a quantum system at the level of {\it a single realization of a sequential measurement -- a trajectory of meter's clicks -- where each readout is related to the position and momentum of the observed quantum particle at the given time}. Although position and momentum are not commuting variables, their simultaneous measurement is possible, though limits on precision are set by the Heisenberg uncertainty relation, as discussed by Arthurs and Kelly \cite{Arthurs65}, and later by Scott and Milburn \cite{Milburn01}. We assume both for Born's rule, stating that the probability of transitioning to a (detector) state is given by its squared overlap with the wavefunction, as well as von Neumann's postulate that after the measurement the system collapses to said "detector state". After such a measurement, the wavefunction does not preserve any information of its pre-measurement state.

Position and momentum are canonical variables of unique importance. They define a phase-space -- the stage for the theory of classical physics with trajectories as the fundamental object. The problem of observing unique trajectories of quantum particles was first raised by Darwin  \cite{Darwin1929}, who in 1929 noticed that though an $\alpha$-particle emitted by a nucleus should be described by a spherical wave, it leaves a straight continuous track in a cloud chamber. The issue was taken by Mott \cite{Mott1929} who formulated the problem as follows: {\it " ... the $\alpha$-particle, once emerged, has particle-like properties, the most striking being the ray tracks that it forms in a Wilson cloud chamber. It is a little difficult to picture how it is that an outgoing spherical wave can produce a straight track ..."}.\\

He resolved the trajectory issue by noticing a proper setting for the problem: accounting not only for the $\alpha$-particle but also for every atom in the Wilson chamber -- the detectors. Only those located at a straight line passing through the nucleus have a significant probability of joint ionization. This marks the first description of emergence of classical trajectories from the wavefunction formalism. In this case, ignoring the back-action of detectors on the highly-energetic $\alpha$-particle was reasonable. At low energies this approximation is not valid - a quantum particle is disturbed by an observation. \\

%To observe a trajectory the continuous monitoring of a particle's position and momentum is needed. 
One peculiarity of standard quantum mechanics is that a probabilistic quantity - the wavefunction - is at the very heart of its formalism.   Comparing theoretical predictions with experimental observations in quantum theory inheres the necessity of repeated measurements. In a vast majority of experiments single-particle observables are measured on an ensemble of identically prepared samples. Results are then averaged over the ensemble - a probabilistic description fits perfectly to such an arrangement.  The situation changed dramatically with advent of ion trapping \cite{Toschek78}. \\

In 1985 Cook and Kimble, based on the idea of Dehmelt \cite{Dehmelt75}, proposed to monitor {\it in situ} quantum jumps - transitions of an electron between levels of a single trapped atom (ion) \cite{Kimble85}. The quantum jumps to or from the  metastable state would be marked by a sudden interruption or resumption of the fluorescence of an atom (ion) resonantly driven at the optical transition. The picture of a jumping electron is present in our thinking about emission or absorption of radiation since Niels Bohr introduced his model of the atom more than a century ago \cite{Bohr1913}. Nevertheless the relation of such discontinuous processes to standard quantum mechanics is a matter of debate (see \cite{Wilkens97} for discussion). \\

Rather than with the Schr\"odinger equation, the rapid switching between bright and dark periods in the fluorescence signal as observed in 1986 \cite{Dehmelt86,Toschek86,Wineland86} seem to be understood in terms of a stochastic process. However, the discontinuities in fluorescence do not necessarily imply classical stochastic electron jumps, as noted in \cite{Juha86}. The stochastic jumps might simply be induced by the measurement of the fluorescence giving a classical random sequence of photon counts. Measurement links the quantum and classical world and it is not clear at which stage, on the part of the atom or the photons, the transition between the two worlds takes place.\\

The continuous quantum mechanical formalism was merged with a stochastic approach in the Monte-Carlo Wave-Function   (MCWF) method introduced by Dalibard, Castin and M\o lmer \cite{Dalibard92}. The method is designed to describe the dynamics of a small system coupled to a Markovian environment. The standard treatment of such a problem involves solving the master equation, which yields the evolution of the density matrix. However, the master equation can be substituted by an ensemble of pure state vectors of the small subsystem where each vector undergoes some random evolution \cite{Dalibard92,Molmer93}. The smooth (non-unitary) evolution of the wavefunction describing the system is interrupted by random quantum jumps due to the interaction with the environment. Each "quantum trajectory" obtained this way can be interpreted as a single realization of the systems' dynamics. Averaging over many such realizations gives a smooth evolution equivalent to the solution of the Gorini-Kossakowski-Sudarshan-Lindblad equation for the density matrix in the Markovian approximation \cite{Gorini76, Lindblad76}. In the following we will use this approach to describe the collapse of the system upon a measurement. A very similar quantum trajectory method was developed by Carmichael \cite{Carmichael92,Carmichael93}. Similar approach based also on a Monte Carlo method was introduced by Dum, Gardiner and Zoller \cite{Dum92,Gardiner92,Dum}. \\

Below  we define a theoretical scheme of repeated measurements on the "same system" in the language of quantum jumps. We use formalism allowing for tracing of a single quantum realization of the process. It is based on the Monte-Carlo Wave-Function   approach. We define the model of detectors  to monitor the particles and discuss its implementation. Our formalism   accounts directly for the von Neumann  (L\"uders) postulate \cite{vonNeumann32, Luders51} that upon measurement the system (not the environment) is found in the state the apparatus is "projecting on", $\phi(x) \rightarrow f(x-x_0)$, where $\phi(x)$ describes a state of the system and $f(x-x_0)$ a meter centered at position $x_0$. By this means we incorporate into the scheme the postulate of wavepacket reduction. Our model of detection conceptually differs from  approaches where detectors act as "filters".  \cite{Barchielli82,Ghirardi86,Caves87,Ueda17}, $\phi(x) \rightarrow \, \propto \phi(x) f(x-x_0)$.  The "measurement operators" we introduce do not correspond to a set of orthogonal states. Therefore our approach is a particular realization of the Positive Operator-Valued  Measure (POVM). We will refer to the scheme we introduce as the "collapse approach".\\

First we illustrate the method in action by showing trajectories of a particle appearing in course of a continued measurement. Then we study in detail a single detector and show how the Zeno effect emerges in the limit of very intense measurement. Next we consider spatially sparse detectors and show that the observed motion of a free particle varies depending on the frequency of observation; from stroboscopic jumps for rarely clicking meters,  up to the Zeno-kind delayed  motion for a very frequent interrogation of the system. We analyse the particle's motion in terms of a stochastic renewal process. Finally  we study the statistical properties of the observed trajectories, i.e. mean values of position as well as dispersion in position and momentum. We show how these quantities scale with time for a dense and a sparse spatial grid of detectors. These results are compared to another model of continuous measurement of position which is based on filtering of the particle's wavefunction in position space \cite{Ueda17}. We summarize the results in the final section.

\section{Method}
Our exemplary object of studies is a particle "observed" by a grid of detectors. The approach can be generalized to a larger number of particles, though this is not the subject of the present study. Each of the detectors is able to measure simultaneously position and momentum of a quantum particle. The detectors are located at discrete points in phase space on a rectangular grid of spacing $D_x \times D_p$. This way positions and momenta measured can take discrete values of $(x_m, k_n)$, and the spacing is $D_x=x_m-x_{m-1}$, $D_p=k_n-k_{n-1}$.  
As we will elaborate at a further point, each possible measurement is associated with a projection on a single detector wavefunction. A convenient choice for the detector wavefunction (and thus of the post-measurement state of the particle) in one dimension (and with trivial generalization to higher dimensions) is the Gaussian wavepacket centered at $(x_m, k_n)$ having a width in position space equal to $\sigma$: 
\begin{equation}
    \langle x|\alpha_{mn}\rangle = \frac{1}{(2 \pi \sigma^2)^\frac{1}{4}} e^{-\frac{(x-x_m)^2}{4 \sigma^2}} e^{i k_n x}.
\label{Dij}
\end{equation}
Every measurement in the phase space with readout values $x=x_m$ and $p=k_n$ is the result of a collapse of the wavefunction on the detector state $|\alpha_{mn}\rangle$.   
In the limit of infinitely small detector spacing, the functions $\alpha_{mn}$ become coherent (squeezed coherent) states of an overcomplete, non-orthogonal basis of the particles' Hilbert space. The detector wavefunction in momentum space is the Fourier transform of (\ref{Dij}), i.e.
\begin{equation}
    \langle k|\alpha_{mn}\rangle = \frac{1}{(2 \pi \sigma^2)^\frac{1}{4}} \sqrt{2 \sigma^2} e^{-\sigma^2 (k-k_n)^2 + i x_m k_n}.
\end{equation}
Evidently, the resolution of measurement in position space is equal to $\sigma_x=\sigma$ while in momentum space to $\sigma_p = 1/2\sigma$. This way the simultaneous measurement of the noncommuting observables minimizes the Heisenberg uncertainty principle \cite{Arthurs65}, 
\begin{equation}
    \sigma_x \sigma_p =\frac{1}{2}.
\end{equation}
Here we work in dimensionless units. In particular both the reduced Planck's constant and the mass of the particle are set to one, $\hbar = m = 1$. We also specify $a_0=\sqrt{2} \sigma$ as our unit of length. In these units the detector wavefunctions have the same width in position and momentum space meaning that the detectors have equal "precision" in measuring both conjugate variables. Similarly, the spacings between detectors in position and momentum are assumed to be equal, $D_x=D_p=D$. The unit of time is $\tau_0= m\sigma^2/\hbar$. Since every detector wavefunction is associated with a measurement outcome $(x_m,k_n)$ our approach differs from the method introduced by Scott and Milburn \cite{Milburn01}, where position and momentum are measured by two principally independent detectors.\\

The crucial issue is to define the effect of the detectors on the particle under observation.  To this end we denote by 
$C_\alpha$ the jump operators acting in the space of the small system, where, to simplify notation, by  $\alpha$ we denote the detector state $\alpha =|\alpha_{m,n}\rangle $.  The  jump operators describe the effect of the environment  on the particle. Their choice is crucial as it defines the physical processes of interest. Here we want to associate the jump with an act of measurement,  \cite{Carmichael92,Carmichael93}.  Therefore we define $C_\alpha$ to be proportional to be projectors onto the $|\alpha\rangle$ state:
\begin{equation}
    \label{Kraus}
    C_{\alpha} = \sqrt{\gamma}|\alpha\rangle\langle \alpha|.
\end{equation}
Here $\gamma$ is a coupling strength determining the characteristic detection frequency. We assume that it has constant value, but in general it can depend both on position and momentum of the meter $\alpha$. This way, we incorporate Born's rule and the von Neumann postulate of wavefunction collapse upon measurement into our approach, \cite{Born1927,vonNeumann32, Luders51, Khrennikov09}. To generate individual sequence of detectors' clicks we make use of the Monte-Carlo Wave-Function (MCWF) method \cite{Dalibard92, Molmer93}. The following paragraphs give a summary of the formalism.\\

\begin{figure}[b]
\includegraphics[width=\linewidth]{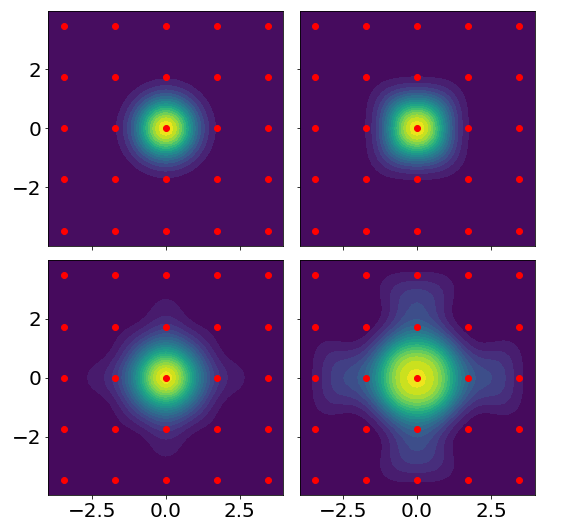}
\caption{\label{fig:evolution} Evolution of the  wavefunction of a particle initially at rest in 2D without quantum jumps. Probability density is plotted. Contrary to the case of a free particle, the particle's wavefunction does not simply broaden, but also tends to concentrate around neighbouring detectors.}
\end{figure}

In detail, the procedure consists of one of two options for each small time step $\delta t$ of the evolution. First the wavefunction $| \phi(t) \rangle$ is evolved using the non-hermitian Hamiltonian:
\begin{equation}
H=H_S- \frac{i}{2} \sum_\alpha C^\dagger_\alpha C_\alpha,
\label{nonhermitian}
\end{equation}
where $H_S$ is the Hamiltonian of the small system, $H_S=\frac{k^2}{2}+V(x)$.
For an infinitely small time increment $\delta t$ we can write: 
\begin{equation} 
\label{eq:time_evolution}
        |\phi^{(1)} (t+ \delta t) \rangle = \left(1-i H \delta t \right) |\phi (t) \rangle.
\end{equation}
The above evolution is non-unitary so the norm is not preserved
\begin{equation}
    \langle \phi^{(1)} (t+ \delta t) |\phi^{(1)} (t+ \delta t) \rangle = 1 - \sum_\alpha \delta p_\alpha,
\end{equation}
with
\begin{equation}
        \label{eq:jump_prob}
        \delta p_\alpha = \delta t \langle \phi (t) | C^\dagger_\alpha C_\alpha | \phi (t) \rangle
\end{equation}
The $\delta p_\alpha$ are interpreted as probabilities of quantum jumps 'in direction' of the respective states $C_\alpha |\phi(t) \rangle$. Thus, the wavefunction at $t+ \delta t$ is selected according to the values of probabilities $\delta p_\alpha$, i.e.: 
\begin{enumerate}
    \item with probability $1-\sum_\alpha \delta p_\alpha$ the wavefunction is the one obtained from nonunitary evolution (with necessary normalization)
    \begin{equation}
    \label{nojump}
        |\phi (t+ \delta t) \rangle = \frac{|\phi^{(1)} (t+ \delta t) \rangle}{||\,|\phi^{(1)} (t+ \delta t) \rangle \,||},
    \end{equation}
    \item one of the meters clicks with probability $\delta p_\alpha/\delta p$ and the particle jumps to the measured state
    \begin{equation}
    \label{jump}
        |\phi (t+ \delta t) \rangle = \frac{C_\alpha |\phi(t) \rangle}{||C_\alpha |\phi(t) \rangle||}=|\alpha\rangle.
    \end{equation}
\end{enumerate}    
Evidently $\delta p=\sum_\alpha \delta p_\alpha$  needs to be smaller than 1. This is usually achieved by tuning the timestep parameter $\delta t$ to be sufficiently small. Note also that although we study the case of a single particle possibly in an external potential, the evolution is far from trivial due to dissipation, i.e. since the particle permanently "feels" all the detectors in its proximity.\\

Assuming that the particle wavefunction itself is roughly Gaussian one can get from Eq.~(\ref{eq:jump_prob}) an estimate for the total jump probability per time step, $ \delta p \approx \gamma \delta t \left( \frac{4 \pi^2}{ D} + 1 \right)$. This formula is particularly useful when choosing appropriate parameters for numerical calculations. It provides a good estimate for both limiting cases of dense (small $D$) and sparse (large $D$) detector grids.\\ 

The evolution of the initial coherent state $\alpha_0$  which is generated by the free Hamiltonian $H_S$ and "dissipation" implied from presence of the meters is illustrated in Fig.~\ref{fig:evolution} for a motionless particle located  at the central detector. In the consecutive panels it is shown that as the time grows (left to right, top to bottom), in addition to quantum diffusion, the modulus (squared) of the wavefunction tends to increase around the positions of the meters -- as if the detectors "attracted" the particle. Note that we presented here a scenario in which \emph{no jumps} occurred within the given time interval. \\

In MCWF, single "trajectories" $| \phi(t) \rangle$ of the system are generated via a stochastic method. Given the same initial conditions, the average $\overline{| \phi(t) \rangle \langle \phi(t) |}$ over multiple trajectories is equal to $\rho_S$, the density operator of the particle, in the limit of many realizations.\\

In the quantum statistical description we treat the particle as a (small) open system coupled to the "reservoir" of detectors. Averaging the individual trajectories one show that the statistical operator satisfies the Gorini-Kossakowski-Sudarshan-Lindblad equation, \cite{Gorini76, Lindblad76} for the (sub)system's density operator $\rho_S$: 
\begin{equation}
    \dot{\rho_S} = i [ \rho_S, H_S ] + \mathcal{L}_{relax} (\rho_S),
    \label{master}
\end{equation}
 and the relaxation operator $\mathcal{L}_{relax}$ describes the interaction with the environment. For general Markovian and time-homogeneous processes $\mathcal{L}_{relax}$ is of the Lindblad form, which for the jump operators defined by Eq.~(\ref{Kraus}),  is:
\begin{equation}
    \mathcal{L}_{relax} = - \frac{1}{2} \sum_\alpha \left(C_\alpha^\dagger C_\alpha \rho_S + \rho_S C_\alpha^\dagger C_\alpha \right) + \sum_\alpha C_\alpha \rho_S C_\alpha^\dagger,
\end{equation}
and summation is extended over the set of all detectors' states, $\sum_\alpha=\sum_{\{\alpha_{mn}\}}$.
Solving (\ref{master}) leads to statistical predictions such as one-time operator average values $\langle A \rangle (t)=Tr(A \rho_S (t))$ within the conventional probabilistic interpretation of quantum mechanics, but it leaves open the question of how the evolution of the system in a single realization might look like.
The MCWF method is thus fully equivalent to the master equation treatment when it comes to predicting statistical quantities. On the other hand in practice it may provide a computational advantage as well as possible additional physical insight from studying the pre-averaged single trajectories. Such observations are well within the reach of modern experimental techniques and we want to include them into our model.\\

We now briefly discuss the relation of this procedure with the stochastic Schr\"odinger equation approach, which is often found in related studies, \cite{Barchielli84,Diosi88a,Diosi88b,Barchielli12}. These approaches usually postulate a stochastic differential equation as a generalization of the standard Schroedinger/von Neumann equation, supplied with some dissipative term, accounting for the continuous observation. Such an equation is mathematically interpreted within the Ito formalism or an alternative quantum stochastic calculus. Indeed, there is a correspondence between the master equation Eq.~(\ref{master}) and various approaches yielding individual realizations, whose expectation values comply with the solution of  Eq.~(\ref{master}) -- the Monte Carlo Wave-Function method being one of them. In our case, the corresponding stochastic equationF would be given as
\begin{equation}
\begin{split}
    d | \psi(t) \rangle &= -i \left( H_S- \frac{i}{2} \sum_\alpha C^\dagger_\alpha C_\alpha + \langle  C^\dagger_\alpha C_\alpha \rangle \right) | \psi \rangle dt \\ 
    & + \sum_\alpha \left( \frac{C_\alpha |\psi \rangle} {||C_\alpha |\psi \rangle || } - |\psi \rangle \right) dN_\alpha(t),
\end{split}
\label{SSE}
\end{equation}
where $dN_\alpha(t)$ is a marked point process \cite{Barchielli12}
and $dN_\alpha(t)dN_\beta(t) = \delta(\alpha-\beta) dN_\alpha(t)$ and $\langle dN_\alpha(t) \rangle = \langle C^\dagger_\alpha C_\alpha\rangle dt$.\\

\section{Exemplary dynamics of a monitored particle}

\subsection{Initial state} 
In principle the initial state may be an arbitrary wavefunction $|\psi(0)\rangle$. However, after the first detection event ("click") this state collapses to a Gaussian wavepacket $\alpha_i$ and all memory of $|\phi (0) \rangle$ is lost. It is therefore of particular importance to analyse the statistics of these first clicks. Note that for jump operators (\ref{Kraus}), the collapse probabilities (\ref{jump}) are given by
\begin{equation}
\begin{split}
    \delta p_{\alpha_i}(t) & = \left(\delta t \gamma \right) \langle \alpha_i | \phi(t) \rangle \langle \phi(t) | \alpha_i \rangle \\
    & = \left(\delta t \gamma \right) \langle \alpha_i |\rho(t) | \alpha_i \rangle \propto Q(\alpha_i,t).
\end{split}
\end{equation}
Assuming very dense detectors, this is essentially (up to a constant) the Husimi representation $Q(\alpha_i)$ of a quantum state. It gives an approximation for the distribution (not normalized to the unity) of the first clicks in our process. We  assume that the first click occurs at $t=0$, at a detector state chosen according to $Q(\alpha)$.

\subsection{Emergence of classical trajectories}
It was shown in many theoretical studies that frequent or continuous observations of a system lead to the emergence of  classical trajectories \cite{Joos85,Zeh85,Rigo97} or even  a classical chaotic motion  \cite{Jacobs00,Graham,Oliveira14}.\\

\begin{figure}[b]
\includegraphics[width=\linewidth]{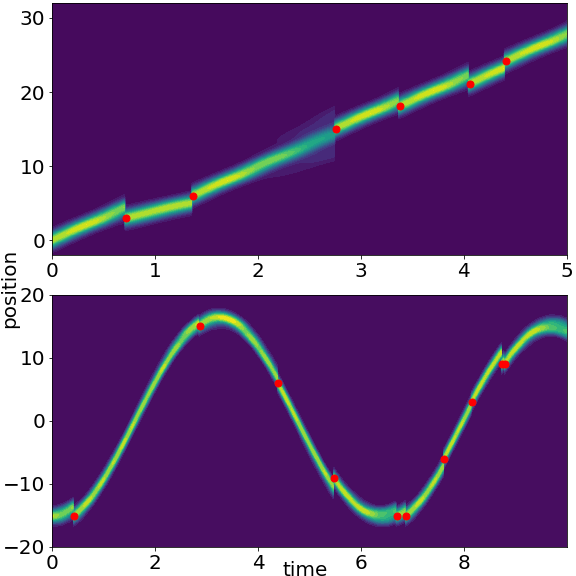}
\caption{\label{fig:trajectory1d} Sample evolution of the probability density $|\phi(x)|^2$ of a particle in free space (upper) and an external harmonic potential (lower panel). In the first case the particle has some nonzero initial momentum, in the second it is located initially at rest at some distance from equilibrium. Detections (marked by red dots) are clearly visible as discontinuities in the vertical direction. Both simulations were performed at $\gamma=2$ and a moderately sparse grid.}
\end{figure}

The method introduced in the previous section provides trajectories as sets of detection events, each consisting of a position and momentum label of the detector activated and a corresponding timestamp. From a numerical point of view, we also have access to the full wavefunction of the particle inbetween detections. However, since only the detection events are supposed to model actual data obtained in an experiment, henceforth we pretend to know nothing more about the evolution of the particle, unless explicitly stated otherwise. As long as detection is sufficiently frequent (with respect to some characteristic time of the system) we assume that this allows us to meaningfully reconstruct the trajectories of a particle and study its properties. We begin with an exemplary trajectory with the spatial separation of detectors chosen to be $D=3$, so that the overlap of two neighbouring meters is rather small ($\sim 0.1$). In the upper panel of Fig.~\ref{fig:trajectory1d} we show the trajectory of a particle with initial momentum $k_0=5$ moving on such a grid of detectors. Red points signify measurement events. Evidently they are aligned along a trajectory, which only slightly deviates from the expected straight line. This deviation is due to the effect of measurement back-action onto the system and the stochastic nature of the process.  In addition to the detected positions we show the probability density as given by the Monte Carlo Wave Function formalism. Collapse upon measurement results in visible discontiuities of the wavefunction in time. Similarly, the lower panel of Fig.~\ref{fig:trajectory1d} shows a sample trajectory of a particle in an external harmonic potential $V(x)=\frac{1}{2}x^2$. The particle is initially positioned at some distance from equilibrium. As expected the oscillatory motion is clearly visible.\\

\begin{figure}[b]
\includegraphics[width=\linewidth]{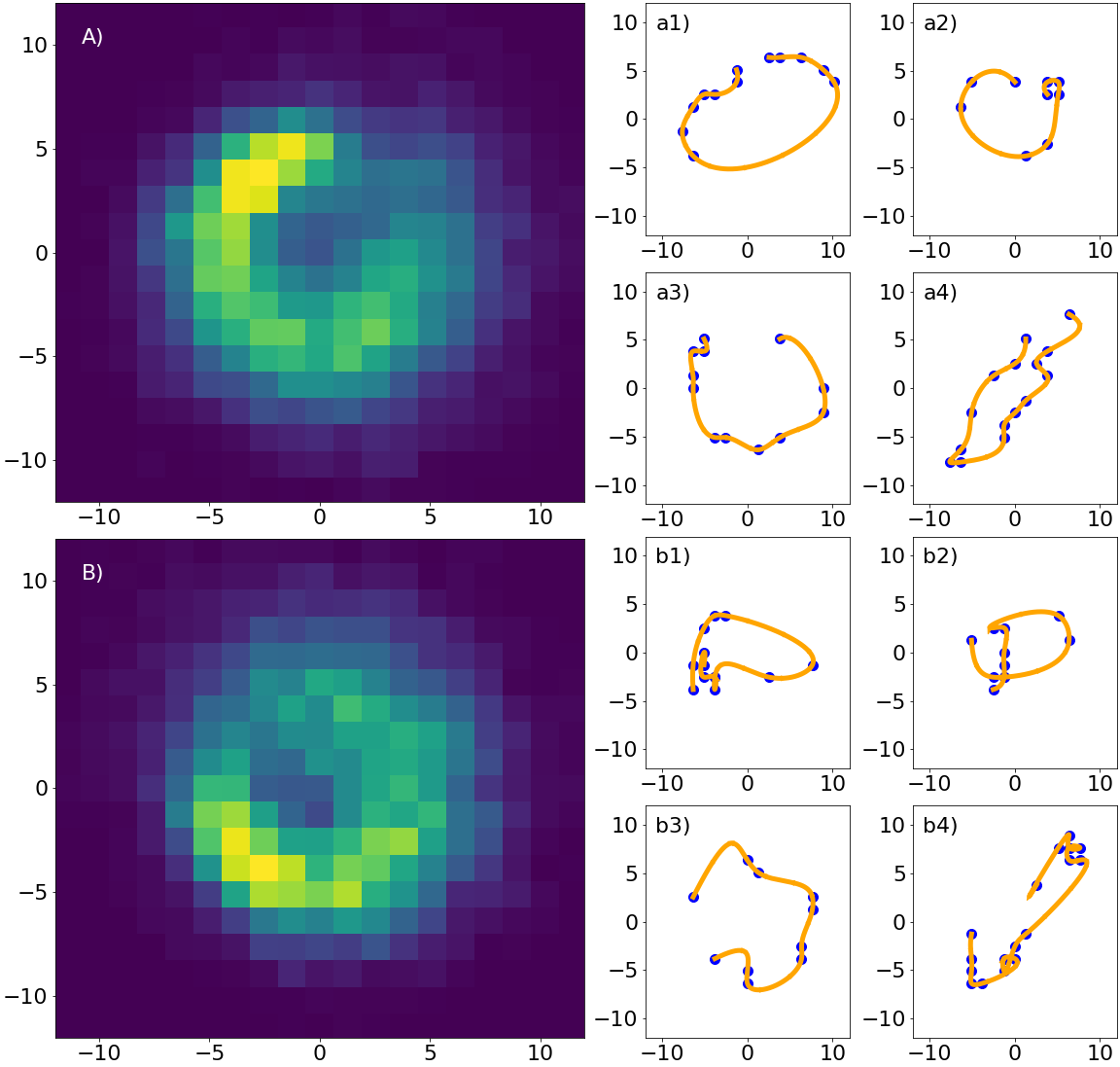}
\caption{\label{fig:trajectory} Left column -- Histogram of trajectories for a particle in a 2D harmonic potential and nonzero angular momentum; A) position space, B) momentum space. For this histogram we simulate trajectories with the first measurement sampled from the distribution given by Eq.~(\ref{circle_psi0}), and post-selected in position in order to only have runs starting in a circular sector covering the upper region of the distribution centered at $(x,y) = (0,5)$, $(k_x,k_y)=(-5,0)$. Right colomn; a1-4) show exemplary individual trajectories in position space; b1-4) show the same trajectories in momentum space.}
\end{figure}

As the final example we present the evolution of a particle in a 2D harmonic oscillator potential $V(r) = \frac{1}{2} \omega (x+y) ^2$ with nonzero initial angular momentum. We choose the initial state in accordance with 
\begin{equation}
    \chi_{l_z}(\rho, \phi) \sim e^{-\omega \rho^2 /2} (\sqrt{\omega} \rho e^{i \phi})^{l_z}
    \label{circle_psi0},
\end{equation}
where $(\rho, \phi)$ are polar coordinates and $l_z=25$ is the quantum number associated with the angular momentum $L_z$. Assuming only the trapping potential and no detectors this state would be stationary, i.e. symmetric under time translation. However, detecting the particle at some instant breaks this symmetry and triggers motion. Upon repeated particle detection, this motion is visible as some form of circular movement. This type of dynamics may be regarded as a caricature of a time crystal, see \cite{Wilczek12}. In  Fig.~\ref{fig:trajectory} we show the evolution as averaged over many quantum trajectories. The top left panel shows the distribution of clicks in position space for many trajectories, integrated over time. In the four top right panels, Fig.~\ref{fig:trajectory}a1-a4, we show examples of individual classical trajectories, i.e. the particle's positions detected at different instants. The orange line is drawn to guide the eye and visualize the time sequence of the different measurements. A  counter-clockwise circular motion as well as perturbations due to interaction with the meters is visible. The bottom panels show analogous plots in momentum space.  In order to be able to visualize the statistical average of the trajectories as seen in the left plots, without loss of generalization we assumed the  the first detection to be at a pre-determined angle $\phi$. This allowed us to account for the rotational symmetry of the state, broken with the first measurement. \\

\subsection{Single detector} 
Quantum systems under frequent observation are known to exhibit a range of phenomena known as the (anti-)Zeno effect \cite{Misra77,Lewenstein00}. This term relates to the fact that the decay of a quantum state may be arrested (or enhanced) under repeatedly performed measurements of a given kind. We now wish to study whether our model (with the measurement operator specified in Eq.~(\ref{Kraus}) exhibits similar features. We will proceed  introducing incrementally more complex configurations of detectors.\\

We begin with the simplistic case of a particle detected at $t=0$ at a single meter centered at position and momentum $\alpha = (x_0=0, k_0)$. We would like to observe the "leakage", i.e. the probability as a function of time with which the wavepacket will escape the reach of the detector. The particle travels with group velocity equal to  the central momentum of its Gaussian profile, $v=k_0$, moving away form the detector. The detector is able to "prevent" this escape by measurement and projecting back onto the meter state. One might expect that if this happened sufficiently frequent, the particle's motion would be effectively frozen, resulting in a typical Zeno effect \cite{Misra77,Gagen93, Fearn92}. In the following we discuss this question in more details.\\

Obviously, for fixed $v$, the entire process will be a function of the $\gamma$ parameter only. Observe first that in accordance with (\ref{eq:jump_prob}), the probability rate of jumping back to the detector state (i.e. the initial state) is given by
\begin{equation}
    \frac{\delta p}{\delta t} = \gamma | \langle \phi (t) | \alpha \rangle |^2,
    \label{rate}
\end{equation}
i.e. is proportional to the modulus (squared) of the overlap of the system wavefunction, $|\psi(t)\rangle = e^{-iHt}|\alpha\rangle$, and the detector state, $|\alpha\rangle$. The probability rate is large when both: i) the particle is located close to the detector's spatial location $x_0 = 0$, and ii) if its velocity $v \propto \frac{1}{2} {\cal I}m\psi \nabla \psi$ matches the momentum of the meter $v \simeq k_0$.  \\

For small time, if  $|\phi (t) \rangle  \approx |\alpha\rangle$, Eq.~(\ref{rate}) simplifies to $\delta p / \delta t = \gamma$. We may thus interpret $\gamma$ as a decay rate: its inverse, $\tau = 1 / \gamma$, gives the characteristic time of the meter's clicks, i.e. the return time to the detector position.\\

We can further quantify the limit mentioned above. The rate with which $|\phi (t)\rangle$ evolves away from its initial state is characterized by two times: First it is $t_1 = \sigma/v \sim 1/v$, which accounts for the particle traveling the distance equal to the sensitivity range of the detector $\sim \sigma$. Second it is $t_2 = 2m \sigma^2 / \hbar \sim 1$, which accounts for quantum dispersion of the wave-packet.  We can identify the regime
\begin{equation}
  \tau << \mathrm{min} (t_1, t_2=1),
   \label{eq:poisson_pp}
\end{equation}
where the particle remains in the reach of the meter. \\

The effect of unitary dynamics  generated by the Hamiltonian $H_S=\frac{k_0^2}{2}$ may be compared against the dissipation due to the coupling to the detectors  $ \frac{i \gamma}{2} | \alpha \rangle \langle \alpha |$.
If $k_0^2 \ll \gamma$ the particle's dynamics is dominated by the non-unitary term - the wavefunction mainly leaks backs to the meter while moving away from the detector with the velocity $v$.
In the limit of infinite $\gamma$, given $|\phi (0) \rangle  = |\alpha \rangle$, we may find the time dependence analytically as
\begin{equation}
|\phi (t) \rangle \propto    e^{-iHt} |\phi (0) \rangle = e^{-\frac{\gamma }{2} t} |\alpha \rangle,
\end{equation}
and if one accounts for the normalization, $|\phi (t) \rangle = |\alpha \rangle $.
Thus, in this limit the detector state is stationary. \\

Monitoring by the detectors manifests itself in two ways. In addition to the described smooth nonunitary dynamics (which occurred to give a stationary probability of the meter clicks), irregular detection events collapse the wavefunction onto the detector state. These are the events seen by the "observer". Since in this case the intensity of the clicks $dp/dt=\gamma$ is constant, the meter provides clicks in accordance with the Poisson point process.  The probability distribution of the time interval between two consecutive measurements is thus an exponential distribution, $e^{-\gamma t}$, and the probability of observing a particle  $n$-times in the time interval $t$ is equal to $P_{capt}(n)=\frac{({\gamma t})^n}{n!}e^{-{\gamma t}}$.  \\

\subsection{Two separated  detectors}

\begin{figure}[b]
\includegraphics[width=\linewidth]{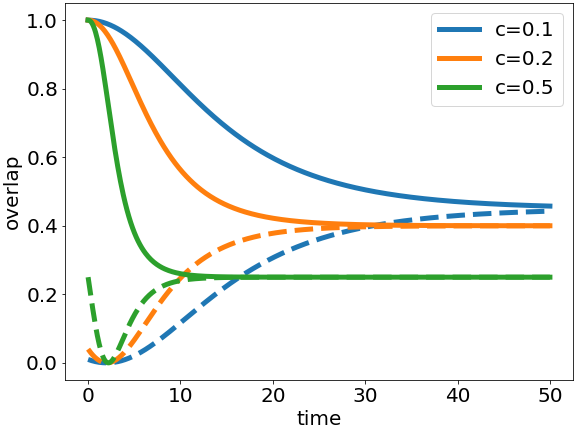}
\caption{\label{fig:two_det} Time dependence of the probability rates $\frac{1}{\gamma} 
\frac{\delta p_\alpha}{{\rm d}t}
=|\langle \alpha |e^{-iHt}|\alpha\rangle|^2 \equiv \frac{1}{\gamma} \lambda_\alpha$ (full lines) and $\frac{1}{\gamma} \frac{\delta p_\beta}{{\rm d}t}=|\langle \beta| e^{-iHt}|\alpha\rangle|^2 \equiv \frac{1}{\gamma} \lambda_\beta$ (dashed lines) of detecting a particle at $\alpha$ or $\beta$ respectively, provided that at $t=0$  the particle is found at $\alpha$ and assuming no clicks afterwards. At a sufficiently large time both rates are equal, $\frac{\delta p_\alpha}{{\rm d}t} \approx 
\frac{\delta p_\beta}{{\rm d}t}$. The probability rates are given by the overlaps (squared) of the detectors' wavefunctions and the state $|\alpha\rangle$ evolved according to the the non-hermitian Hamiltonian. These quantities define the intensities of the renewal processes $\lambda_\alpha$ and $\lambda_\beta$ corresponding to recapturing of the particle by the detector $\alpha$ and $\beta$ respectively.}
\end{figure}

We now investigate the case of two detectors in the aforementioned limit of large $\gamma$, assuming some variable distance between them. The Hamiltonian is then assumed to be
\begin{equation}
\label{H_nh}
        H = - \frac{i \gamma}{2} \biggr( | \alpha \rangle \langle \alpha | +| \beta \rangle \langle \beta | \biggl),
\end{equation}
Let us assume that these two detectors are centered at $\alpha = (x=0, k_0=0)$ and $\beta=(x=D,k_0=0)$. The purpose of this exactly solvable example is to give some intuition of the dynamics of a slow particle, very intensively observed by a  sparsely distributed meters.\\

Not accounting for jumps, the initial state localized at the meter $\alpha$ evolves as:
\begin{equation}
    e^{-iHt} |\alpha \rangle = e^{-\frac{\gamma t}{2}}  \cosh{\frac{c \gamma t}{2}} |\alpha \rangle -\frac{c^*}{|c|} e^{-\frac{\gamma t}{2}} \sinh{\frac{c \gamma t}{2}} |\beta \rangle 
\end{equation}
The probability rates  $\lambda_\alpha=\frac{\delta p_\alpha}{\delta t}$ and $\lambda_\beta=\frac{\delta p_\beta}{\delta t}$ of a detection at one of the two meters $\alpha$ and $\beta$ can be easily found:
\begin{equation}
\label{d_alpha}
    \lambda_\alpha=\gamma \frac{|\langle \alpha| e^{-iHt} |\alpha \rangle|^2}{|| e^{-iHt} |\alpha \rangle\,||^2} = \gamma \frac{ \left(\cosh\frac{t}{t_0}  -|c| \sinh{\frac{t}{t_0}} \right)^2}{\cosh \frac{2t}{t_0}-\sinh\frac{2t}{t_0}}, 
\end{equation}
\begin{equation}
\label{d_beta}
    \lambda_\beta=\gamma \frac{|\langle \beta| e^{-iHt} |\alpha \rangle|^2}{|| e^{-iHt} |\alpha \rangle\,||^2} = \gamma  \frac{  \left(|c| \cosh\frac{t}{t_0}  - \sinh{\frac{t}{t_0}} \right)^2}{\cosh \frac{2t}{t_0}-\sinh\frac{2t}{t_0}}, 
\end{equation}
where  $c = \langle \alpha | \beta \rangle$ and we identify a characteristic time scale $t_0=\frac{2}{\gamma |c|}$. For sparsely distributed detectors $c\sim \exp({-D^2/4})$ is very small and $t_0 \sim (2/\gamma) \exp({D^2/4}) \gg 1/\gamma$ is very large.\\

Eqs.~(\ref{d_alpha}), (\ref{d_beta}) give some insight into   physical picture of events. First, for time $t<1/\gamma$ the probability to  monitor the particle  at the initial  position of the detector $\alpha$, is large $\lambda_\alpha(t)\delta t  \gg \lambda_\beta(t)\delta t  \sim \gamma |c|^2 \delta t$. The particle is frequently measured only at $\alpha$. Finally an attempt occurs where there is no click within an interval of order $t_0$. $\lambda_\alpha(t) \delta t$ and $\lambda_\beta(t) \delta t$ evolve to grow comparable. This eventually allows the particle to be located at $\beta$, and the whole situation repeats with the role of detectors $\beta$ and $\alpha$ interchanged. Given that there exists some characteristic time of passing from $\alpha$ to $\beta$, the dynamics may be understood as an alternating renewal processes. Contrary to a single detector, when the particle stays frozen at the detector position, the particle has a significant chance to be found at the second meter after sufficiently long time (tending to infinity with $\gamma$). In many respects the process resembles  the tunneling of a particle between two separated potential wells perturbed by observation \cite{Gagen93, Fearn92}.\\

The results presented here may be generalized to the case of a slowly moving particle $v < D/\gamma$ in the limit of very frequent interrogation $\gamma \gg v^2$. The particle  will jump randomly between neighbouring detectors in a kind of diffusive motion (in space and momentum). A quantitative description of the diffusion will be presented in Sec.~(\ref{sec:comparison}).\\

We want to stress that excluding very dense detectors, the transition time from $\alpha$  to $\beta$ is generally significantly larger than $t_0$, and it constitutes a rather weak lower bound. This is because, as we have pointed out, the particle would return many times to $|\alpha \rangle$ before reaching a state of comparable $\lambda_\alpha$ and $\lambda_\beta$. The dynamics of a system randomly recurring to its initial state is known as a renewal process. We elaborate on this in the following section.  \\

\section{Renewal  dynamics}
\subsection{General considerations}

\begin{figure}[b]
\includegraphics[width=\linewidth]{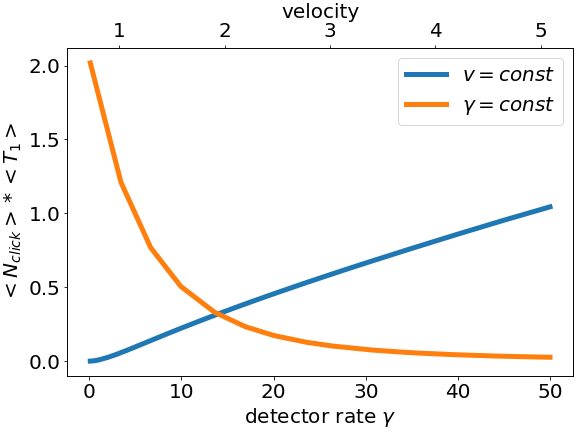}
\caption{\label{fig:escape} The escape time. The blue line (the scale at the bottom edge of figure) shows dependence of the escape time as a function of the detector rate $\gamma$ for a fixed velocity $v=2$. The escape time grows almost linearly with $\gamma$. The larger $\gamma$ the more time it takes to escape. The orange line (the scale at the top edge of the figure) displays dependence of the escape time on the particle velocity for fixed detector rate $\gamma=15$. The larger the velocity, the shorter the time of escape. }
\end{figure}

For a better understanding of the Zeno-like dynamics we observe in certain settings, we now discuss the problem of a particle trying to "escape" a single detector while dropping the assumption of large $\gamma$. The particle's initial position and momentum (chosen to be nonzero) are set to be equal to the detector's eigenstate. The probability of capturing the particle will vary in time and every detection returns the whole system to its initial state. This kind of behaviour is known as a renewal process and is usually difficult to treat analytically. In the following we present a general treatment supplemented by some exact formulas.\\

The inter-measurement evolution is now an inhomogenouos Poisson process (IPP) with intensity function
\begin{equation}
    \lambda (t) =  \frac{\delta p}{\delta t}
\end{equation}

\begin{figure}[b]
\includegraphics[width=\linewidth]{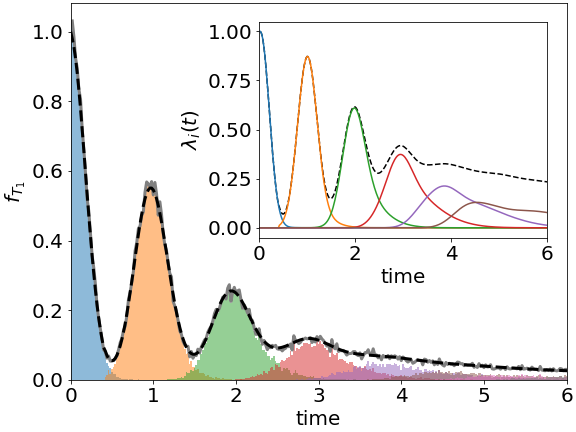}
\caption{\label{fig:inter_arrival} Probability distribution of the arrival time of the first click $f_{T_1}(t)$ for $D=5$, $\gamma=1$ and $v=5$. The colored areas are suitably normalized histograms, with different colors indicating a first click at the n-th detector. The grey noisy line is the sum of these and gives the distribution $f_{T_1}(t)$ of being caught first at an arbitrary detector after time $t$. This line coincides with the black dotted line, which is the same function obtained numerically from Eq.~(\ref{ft1}). Because $\gamma$ is not too large, the positions of the first few colored maxima correspond roughly to the time when the particle would arrive classically at the position of the meters. The data for the histogram was obtained from 100 000 single trajectories and grouped into bins of width $\delta t = 0.02$.  The first 3 or 4 maxima are well separated, which corresponds to visible steps in Fig.~\ref{fig:schodki}. The inset shows the individual intensities  $\lambda_i(t)$ plotted in different colors. Their sum, the total intensity, is depicted by the dashed line. }
\end{figure}

\begin{figure}[b]
\includegraphics[width=\linewidth]{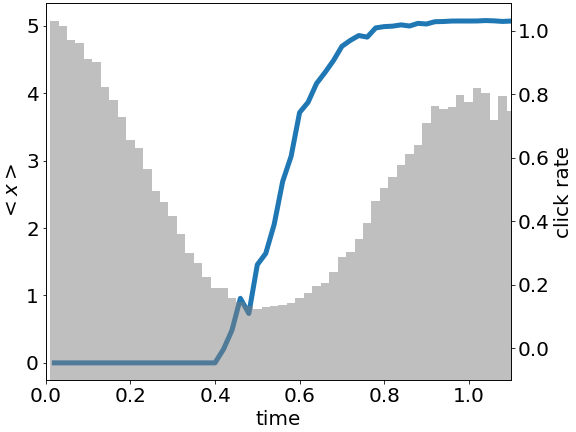}
\caption{\label{fig:steps_hist} Blue line -- the mean position of the particle as a function of time observed by two well-seperated detectors characterized by $\gamma=1$. The  first detector is located at $\alpha_1= (x=0,v=5)$ and the second at $\alpha_2=(x=5,v=5)$. The grey (normalized) histogram shows the frequency of clicks at both detectors (bin width $\delta t= 0.02$). At time $t < 0.5$ the particle is almost exclusively captured by $\alpha_1$ and is thus measured at the position $x=0$.  At larger times $t>0.5$ a growing number of trajectories arrives at the second detector. There is a time window around  $t=0.5$ when the frequency of clicks at both detectors is significantly lower. This is because at this time, the particle tends on average not to have a large overlap with either of the detectors. It is detected at $x=0$ or at $x=5$ with roughly the same small probability, which is why its average position is between the detectors.}
\end{figure}

In this analysis we assume that the function $\lambda(t)$ is well-behaved, in particular that its integral over any interval is finite. Furthermore, an obvious necessary condition for the particle to be able to "escape" the detector is $\lim_{t \rightarrow \infty} \lambda(t) = 0$, which is very plausible for this physical setting (note however that this need not be true for a general arrangement of detectors). A standard result for the inhomogeneous Poisson process is that the probability of zero events on a given interval $[0,t]$ is:
\begin{equation}
    P(N=0) = e^{-\Lambda(t)}
    \label{PN0}
\end{equation}
where
\begin{equation}
    \Lambda (t) = \int_0^t \lambda(t') dt'.
\end{equation}
Thus the probability of the particle escaping the reach of the detector without being captured even once is obtained by calculating the limit of (\ref{PN0}) for infinite $t$. To calculate the probability of a single capture followed by escape, we assume a large interval $[0,T]$ and capture time $t'$ localized within a small intervall $dt'$. In accordance with the above remarks, the probability of such a realization is given by
\begin{equation}
    e^{-\Lambda (t')} \lambda(t') dt' e^{-\Lambda (T-t')}
\end{equation}
This formula may be understood by dividing the timeline into three intervals: before the measurement, around detection, and afterwards. Since we want to account for a detection at an arbitrary instant, we integrate the above equation for $t'$ over $[0,T]$. Since $T$ is supposed to be very large, we approximate $\exp(-\Lambda(T-t') = \exp(-\Lambda(T))$. This finally gives us
\begin{equation}
    P(N=1) = e^{-\Lambda(T)} (1 -e^{-\Lambda(T)}).
\end{equation}
To obtain a formula for escape after two detections, we apply an analogous reasoning, assuming the timeline $[0,T]$ and detection times $t'$ and $t''$. Without loss of generality we assume that $t'<t''$, which means that $t'$ is located anywhere on the timeline and $t''$ within $[t', T]$. This gives us the limits of the corresponding double integral, yielding
\begin{equation}
     P(N=2) = e^{-\Lambda(T)} (1 -e^{-\Lambda(T)})^2
\end{equation}
For the general case we obtain 
\begin{equation}
p(N=n) = p(1-p)^n.
\label{pNn}
\end{equation}
with $p=\exp(-\Lambda(T))$. Eq.~(\ref{pNn}) is the geometric distribution. Its expectation value
\begin{equation}
\langle N_{click} \rangle =e^{\Lambda(T)}-1,
\label{EN}
\end{equation}
is the mean number of detections before the particle escapes the meter.  The inter-arrival time, i.e. the time  between consecutive returns to the detector position due to the collapse caused by the measurement, coincides with the random variable $T_1$ of the time of the first click. In the case of the inhomogeneous Poisson process its distribution is known to be
\begin{equation}
    f_{T_1}(t) = \lambda(t) e^{-\Lambda(t)}
    \label{ft1}
\end{equation}
which allows us to calculate the average waiting time between two clicks as
\begin{equation}
    \langle T_1 \rangle = \int_0^{\infty} t  f_{T_1}(t) dt.
    \label{ET1}
\end{equation}
Equations Eq.~(\ref{EN}) and Eq.~(\ref{ET1}) allows to estimate the escape time, $T_{esc}$, i.e. the  time needed for the particle located at the position of the meter to  escape  its reach. This estimation is given by the number of renewals multiplied by the inter-arrival time, $T_{esc}=\langle N_{click}\rangle \langle T_1 \rangle$:
\begin{equation}
    T_{esc}=\left( e^{\Lambda(T)}-1\right) \int_0^{\infty} t  \lambda(t) e^{-\Lambda(t)} dt
    \label{T_esc}
\end{equation}
In Fig.~\ref{fig:escape} we show the dependence of the escape time on the detector rate $\gamma$ and velocity $v$. The data points were obtained by calculating the respective $\lambda(t)$ numerically and then substituting into Eq.~(\ref{T_esc}). The blue line (lower scale) shows the dependence on $\gamma$: The larger the parameter, the more time it takes to escape. The dependence is close to a linear one. The orange line depicts the escape time for different values of velocity of the particle. The escape time scales approximately as $T_{esc} \sim \frac{1}{v^\kappa}$ where $\kappa$ takes a value between 2 and 3. The larger the velocity the shorter the escape time. \\

\subsection{Rarely interrogated particle}

Equipped with the basic results of the previous section we now move on to analyse the 1D motion of a particle with initial finite velocity $v$ observed by a lattice of detectors of finite  $\gamma$ where both the unitary and the nonunitary evolution compete. We will present some analytical predictions and compare them with numerical results.\\

We start by considering the case of sparsely distributed  detectors, $D \gg 1$, characterized by "low"  clicking rate $\gamma \lesssim 1$.  We may now associate each detector with an intensity function $\lambda_i(t)=\gamma |\langle \psi(t)|\alpha_i\rangle|^2$, where $\psi(0)=\alpha_0$. The sum of these is the total intensity function 
$\lambda(t)$ with the usual relation $\Lambda(t)=\int_0^t \lambda(t') dt'$. Given these definitions the probability distribution of the first clicks (inter-arrival time) $f_{T_1}(t)$ at any detector is again given by Eq.~(\ref{ft1}).\\

To get an approximate expression for $f_{T_1}(t)$ let us observe that for short times after a detection ($t \sim 1$) the evolution may be very well approximated by assuming that for the first $i$ detectors $\lambda_i(t) \approx \gamma \exp{(-[vt-iD]^2)}$. This means essentially neglecting wavepacket spreading due to quantum dissipation and the "leakage" to nearby detectors due to the non-hermitian part of the evolution. 
Similarly, if $\gamma \lessapprox 1$ then $\Lambda(t) \approx \gamma t$, and the probability distribution of the inter-arrival  time for sparsely distributed detectors $D> 1$, small clicking intensity $\gamma \lessapprox 1$, and  $t \lessapprox 1$,  is approximately equal to:
\begin{equation}
\label{fT1}
f_{T_1}(t) \approx \gamma e^{-\gamma t} \sum_{i=0} e^{-[vt-iD]^2}
\end{equation}
The approximation works because the expected number of clicks for this time interval is rather small and breaks for larger $\gamma$, as shown later on. \\

In Fig.~\ref{fig:inter_arrival} we present the probability distribution of the inter-arrival times.  The function $f_{T_1}(t)$ obtained by numerically solving the integral (\ref{ft1}) (black dotted line) perfectly agrees with the histograms obtained from an ensemble of individual trajectories generated using the MCWF method (grey line), see Fig.~\ref{fig:inter_arrival}.  Only small stochastic fluctuations of the quantum trajectories allow to distinguish the two approaches. In this scenario, the particle is initially located at the detector $\alpha_0$ and has velocity $v=k_0=5$. The separation between detectors is $D=5$, therefore $D\gamma/v=1$. This means that in a characteristic interval between consecutive clicks the particle travels roughly the distance equal to the separation between neighbouring detectors. This is close to the edge of the parametric region which can be termed as the "small $\gamma$ region" (cf. the following subsection).\\

The results shown in Fig.~\ref{fig:inter_arrival} can be understood in terms of Eq.~(\ref{fT1}). The positions of the peaks at $t = i \frac{D}{v}$ correspond to the local maxima of the Gaussian functions $e^{-i[vt-iD]^2}$ in $\lambda_i(t)$. The prominence of these peaks depends on the values of $v$ and $\gamma$. For $\gamma/v \lessapprox 1$ the first few peaks are of comparable height. For large $\gamma/v$ the mean of the distribution shifts to the left, so that ultimately only the peak at $t=0$ is prominent ant the others are strongly suppressed. In either case, further peaks are not as prominent, since after a long time without detection, the particle wavefunction tends to delocalize heavily due to the non-hermitian part of the Hamiltonian. However, this tail of $f_{T_1}(t)$ does not play a significant role in the evolution, since such large periods of non-detection happen very seldom. This fact can be utilized when constructing numerical schemes for approximate solution of the dynamics (cf. Fig.~\ref{fig:schodki}) and the following subsection). The accuracy of such a scheme is arbitrary high, tunable by setting a cut-off time $t_c$ such that $\int_0^{t_c}f_{T_1}(t')dt'$ is sufficiently close to 1. The colored areas mark suitably normalized histograms of first clicks at the n-th detector obtained from MCWF simulations $(n=0, 1, \ldots)$ with each color representing a different meter. Their sum gives $f_{T_1}(t)$. Evidently for times $0< t< D/2v$ the first clicks result from the action of the first meter. If the first click occurs at a later time, $D/2v < t < 3D/2v$, it is the action of the second detector, and so on. The inset of Fig.~\ref{fig:inter_arrival} shows the intensities $\lambda_i(t) = \gamma |\langle \alpha_0(t)|\alpha _i\rangle|^2$ obtained by numerical  propagation of the initial state $|\alpha _0(t)\rangle=e^{-iHt} |\alpha _0\rangle$,  according to the nonhermitian Hamiltonian $H$.\\

Choosing the above set of parameters, Fig.~\ref{fig:steps_hist} shows the expected position of the particle as a function of time, calculated by averaging over many trajectories obtained via MCWF (blue line). The grey histogram in the background shows the frequency of detection as a function of time. Contrary to a classic straight line, the function is rather a "step": Around $t=0$ ($x=0$), the particle tends to be (relatively) frequently measured by the first detector. Around $t=0.5$, the detection rate is very low. Finally, as we approach $t=1$ ($x=5$), the detection rate rises again, this time (predominantly) at the second detector. The dispersion of the position measurements is minimal in the vicinity of the plateaus $t=0$ and $t=1$ and reaches a local maximum inbetween (cf. Fig.~\ref{fig:my_label}). Since we find ourselves in the regime of low $\gamma$, the shape of the step is well approximated by the expression  $\langle x(t)\rangle = D \sum_i i \lambda_i (t)/ \sum_i \lambda_i (t)$.\\

\subsection{Frequently clicking detectors}

\begin{figure}[b]
\includegraphics[width=\linewidth]{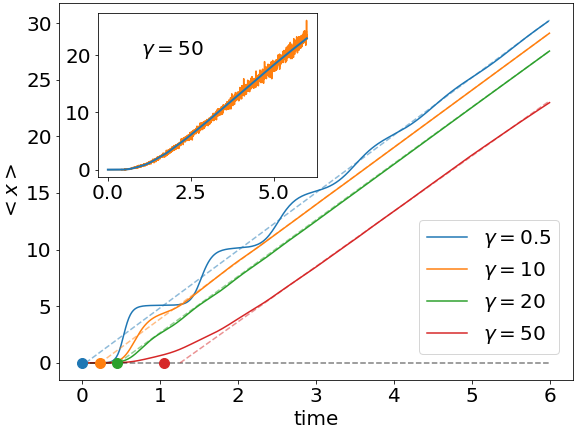}
\caption{\label{fig:schodki} Mean positions as a function of time for a sparse grid of detectors, $v_0 \sim 5$, and various $\gamma$. Note how for large $\gamma$ there is a retardation effect. The retardation $t_r$  may well be approximated by Eq.~(\ref{T_esc}). Estimates from a linear asymptotic fit for $\gamma = 10, 20, 50$: $t_r = 0.24, 0.51, 1.26$ indicated by the colored dots. These curves were obtained using an exact numerical procedure under the assumption that there are no jumps to neighbouring momentum states. This assumption is valid for a sparse grid and was tested with results obtained from a full MCWF simulation - in the extreme case, i.e. for $\gamma=50$, for $N=10^6$ detections only around 0.1\% of all detection events resulted in a change of momentum. A comparison of results from the numerical procedure and the MCWF calculation for $\gamma=50$ is shown in the inset.}
\end{figure}

We now turn to study differences between low and high monitoring frequency $\gamma$. Furthermore we assume that detectors are sparsely distributed, since this is a regime with a broad range of dynamical behaviours. Simulations with large $\gamma$ are computationally demanding since they require to choose a very small timestep $\delta t$. To study statistical characteristics of the dynamics, in addition to the approach based on simulating of individual trajectories, we therefore employ a different method, too.\\

We solve the Gorini-Kossakowski-Sudarshan-Lindblad equation numerically for a 1D geometry by finding the first few jump amplitudes $\lambda_i (t) \propto |\langle \alpha_i|e^{-iH t}|\alpha_0|\rangle |^2$ as a function of time. To simplify the problem we also assume that jumps to states with a different than initial velocity are very unlikely. For a particle with initial velocity $v=5$, detectors spaced by $D=5$ and measurement intensity $\gamma=50$, among $10^6$ simulated quantum trajectories only $10^3$ showed a collapse to a neighbouring momentum state. Using the discrete symmetry of the lattice and the "renewal" property of the dynamics we then employ a numerical scheme to find the clicking rates at the different detectors as a function of time. In the inset of Fig.~\ref{fig:schodki} we show the average position of the particle as a function of time obtained with the MCWF metod (orange noisy curve), while the blue line represents the calculation based on the non-stochastic approach. The agreement is evident.\\ 

The averaged trajectories for different values of $\gamma$ are presented in the main panel of Fig.~\ref{fig:schodki}. We show results for $\gamma=0.5,10,20,50$, starting from detectors with relatively low measurement intensity up to very high values. One can see how the previously described initial stroboscopic motion, characteristic for small $\gamma=0.5$, disappears as $\gamma$ grows. The disappearance of the "steps" is directly related to the more frequent interrogation of the particle by the meters. As detection events become more frequent, the different trajectories in the ensemble get "out-of-sync" earlier, so that comparing several trajectories, different detectors may be active in the same time intervals. We want to stress that the same effects are responsible for smearing out the steps at later times of evolution for small $\gamma$.\\

Interestingly, for large times the average evolution $\langle x \rangle (t)$ is always a straight line with a slope equal to the initial velocity, regardless of the character of the dynamics at the initial stages. By extending the "asymptotic" behaviour (at large time) to the initial position $x=0$ we see that from the point of view of a distant observer it looks like (the ensembles of) the trajectories with different $\gamma$ started at different moments, delayed with respect to the classic case (or the limit $\gamma \rightarrow 0$). This is visualized as the crossing point of the colored dashed lines with the x-axis marked as a black line in Fig.~\ref{fig:schodki}. The delay  grows with $\gamma$ and is reminiscent of a Zeno effect. The value of the delay agrees very well with the formula Eq.~(\ref{T_esc}), which was earlier used to estimate the escape time from a single detector. The numerical values obtained from this equation are shown as colored dots. \\

Clearly, the delay effect stems from the interaction with the first detector, at whose position the particle is initialized. The other meters detecting the particle may just as well accelerate the particle's motion (by detecting it before it would classically arrive at their position).  Because the trajectories quickly grow out of sync clicks, the cumulative effect of retarding/accelerating the particle averages to zero.  \\

Finally, we shortly comment the case of a dense detector grid. A particle moving within such a grid will have at any given time a substantial chance of being measured by many different detectors. The renewal-type dynamics disappear, along with the delay effect and step-like stroboscopic motion for small $\gamma$. The particle follows effectively a classical trajectory with diffusion characterized by $\gamma$ and the grid's density.\\

\section{Comparison with observations based on spatial filtering}
\label{sec:comparison}

We have shown that repeated measurements reveal classical dynamics perturbed by the back-action of the meters, as visible in the exemplary trajectories  Figs.~(\ref{fig:trajectory1d},\ref{fig:trajectory}). Due to the probabilistic nature of "detection" the trajectories form a random sequence of clicks. In this section we present some statistical characteristics of an ensemble of such trajectories. \\

We will compare them with statistics obtained from an alternative model of continouos observation based on measurement described as a spatial filtering of the particle's wavefunction by the filter function centered at $X$ \cite{Barchielli82,Caves87,Ghirardi86,Ueda17}. In these approaches the assumed form of the measurement (Kraus) operator is:  
\begin{equation}
    K(X) = \sqrt{\gamma} \int f(X-x') |x' \rangle \langle x' | dx',
    \label{M_Ueda_b}
\end{equation}
Where $f$ is a suitable localized function. The action of $K(X)$ effectively reduces to multiplying the wavefunction in position space by an envelope centered around the measurement outcome $X$. We are going to refer to the measurement described by the operator $K(X)$ as "filtering", as opposed to the "projection" on coherent states related to our jump operators $C_\alpha$. The filtering approach can be easily incorporated into the scheme of the MCWF method we use here. It amounts to a substitution of the jump operators $C_\alpha$ by the filters $K(X)$.  To have a direct correspondence between the two methods  we assume that in both cases the meters are located on the same spatial grid of spacing $D$. Moreover, we choose the filtering functions $f(x_m-x)$, centered  at $x_m$ to be identical (up to the phase factor $e^{-ix k_n}$) to the Gaussian function of the jump operators  $C_\alpha$, i.e.  we choose $f(x_m-x)=\frac{1}{(2 \pi \sigma^2)^\frac{1}{4}} e^{-\frac{(x_m-x)^2}{4 \sigma^2}}$ with $\sigma = 1/\sqrt{2}$. An important difference between the two approaches is that the effect of filtering results only in a measurement of the particle's position, and not its momentum.\\

Note also that repeated applying the operator $K(X)$ differs from a single measurement. This is in contrast to our method in which the jump operator $C_\alpha$ is proportional to a projection operator. Hence immediately after detection, the particle is in an eigenstate of the measurement operation, with maximum probability of being captured again at this position and momentum (indeed, for an orthogonal set of detector functions this would be the only possible measurement outcome at this point in time).\\

We limit our study to the case of a free particle launched  with initial momentum $k_0=5$ and meters characterized by measurement intensity $\gamma=1$. We assume that at $t=0$ the particle's wavefunction is described by a Gaussian wavepacket located at $x_0=0$ and identical to the spatial profile of the detector $\langle x|\alpha\rangle$ which we set at the origin of the coordinate system, $\alpha=(x_0=0,k_0=5)$. \\

As discussed earlier, the dynamics crucially depend on the detector spacing $D$. According to our choice the separation between different momentum states is also equal to $D$. We shall discuss the case when the spatial distance  between the detectors is larger then the range of their sensitivity, $D=5.1 > \sigma=\frac{1}{\sqrt{2}}$, and the opposite regime  when the detectors cover "almost continuously" the entire space, $D=0.73 \simeq \sigma=\frac{1}{\sqrt{2}}$. In the first case the classical time to travel the distance $D$ is of the order of $t=1$, i.e. it is of the order of the average time between two clicks of a single detector. For the dense grid, the particle arrives at a neighbouring meter in a time significantly smaller than the "reaction" time $1/\gamma = 1$ of the meters.\\

To compare the statistics of trajectories provided by the two models we simulate a number of individual realizations using the MCWF method. In the case of our jump model, each trajectory is a set of measurement events $\{x_i,k_i,t_i\}$. In the case of filtering measurement operators (\ref{M_Ueda_b}), the results only comprise position measurements $x_i$ at times $t_i$, so we infer the particle's velocity at these times as $k_i = \frac{x_{i}-x_{i-1}}{t_{i}-t_{i-1}}$, where $x_0$ is the particle's initial position. We collect these results for both a very dense and very sparse detector grid. As before, the detection events of the trajectory ensemble are grouped into small time intervals, for which we calculate the mean value and standard deviation of $x$ and $k$.\\

\begin{figure}
    \centering
    \includegraphics[width=\linewidth]{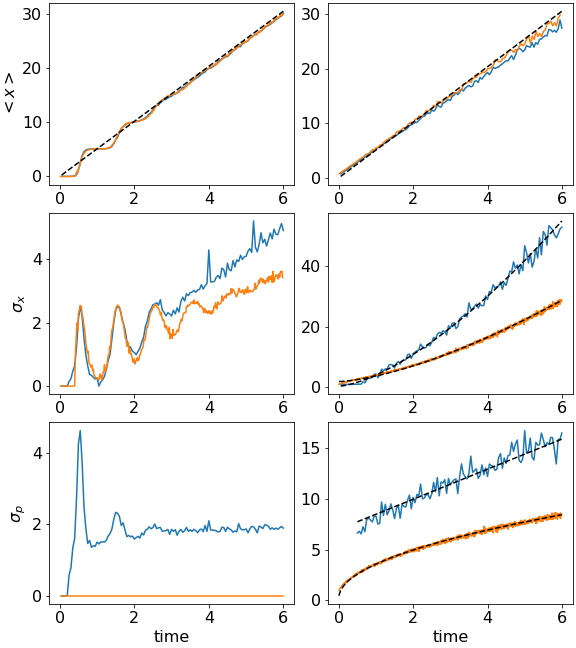}
    \caption{Average position (top) and the dispersion of position (middle) and momentum (bottom) measurements as a function of time for the particle launched with initial velocity $k_0=5$ observed by detectors characterized by $\gamma=1$ and distributed along a line with spacing  $D=5.1$ (left panels) and $D=0.73$ (right panels). The orange line corresponds to the jump approach from the present study. Results from the filtering method are shown for comparison as a blue line. The black dotted lines indicate classical trajectories (top row), quadratic $\sim t^{2}$ (blue) and $\sim t^{3/2}$ (orange) fits (middle row), and linear $\sim t$ (blue) and $\sim t^{1/2}$ (orange) fits (bottom row).}
    \label{fig:my_label}
\end{figure}

The time dependence of the mean position of the particle is shown in the upper left panel of Fig.~\ref{fig:my_label}. In all figures the orange lines correspond to our quantum jump approach while blue lines indicate a filtering measurement. 
As we have already shown,  the initial evolution of the particle localized initially at one of the meters has a stroboscopic (step-like) character if detectors are very sparse. The particle spends some substantial time at one of the detectors and then rapidly jumps to the neighbouring one. As soon as the characteristic width of the wavepacket is similar to the detectors spacing, the observed motion of the particle becomes classical (top right panel). Both approaches give here the same results. The dashed black line shows the classical trajectory of a particle moving with initial velocity $k_0=5$.\\

In the middle panel we present the dispersion of the position of the particle as a function of time. For sparse detectors (left middle panels) both methods predict  large oscillations. They are directly related to the "stroboscopic motion" of the particle predicted at the early times of the evolution. The dispersion has peaks at times classically corresponding to half the distance between the detectors. This effect is analogous to shot noise found in many physical settings. At larger times, the dispersion grows linearly, though the slope is larger in the filtering approach. For dense detector spacing the dispersion of the particle's position scales as $\sim t^{3/2}$ for the quantum jump approach while scaling as $\sim t^2$ for the filtering method (right middle panel). The growth of the dispersion is again faster in the filtering scenario.\\

Significant differences may be observed in the case of the dispersion of the particle's momentum. For sparse detectors, the jump measurement leaves the particle's average momentum constant and its dispersion is negligible (left low panel). This is because for the chosen spacing, it is extremely unlikely to jump to a neighbouring momentum state. In the case of filtering meters the dispersion of the momentum exhibits oscillations at initial time mimicking the change of the particle's average position, and stays constant later on. \\

In the case of dense detector spacing the filtering method gives linear growth of the momentum dispersion, $\sigma_k \sim t$, while the dispersion obtained from the jump measurements grows as $\sigma_k \sim \sqrt{t}$.  This dependence revokes the characteristic of a random walk in momentum space. This is supported by the fact that the momenta maintain a Gaussian distribution around their mean (i.e. initial) value over time. \\

From a theoretical point of view this scaling of the dispersion in the quantum jump scheme is plausible since indeed, in the absence of an external potential the particle always has the same probability of jumping to $k+dk$ as to $k-dk$. In the limit of dense detectors and frequent measurement we thus expect that the random walk will pass into a Wiener process, and the momentum distribution will be governed by a standard diffusion equation, which is physically equivalent to constantly "heating" the system.\\

If we do not wish the system to "heat up" indefinitely, the approach must be generalized. The simplest way is  to introduce a dependency of the coupling strength $\gamma$ on the absolute momentum measured by the respective detector $\alpha_{m,n} =(x_m, k_n)$. One particularly simple approach is to set $C_{\alpha_{m,n}}=\sqrt{\gamma_n}|\alpha_{m,n}\rangle\langle\alpha_{m,n}|$ where 
$\gamma_{n}=\gamma_0 e^{-k_n^2 / k_{cut}^2}$,
where $k_{cut}$ is a characteristic cut-off momentum above which the detectors are supposed to become less sensible. One might also consider different modifications of the scheme going beyond the Markov approximation. This is however not the issue of the present study.

\section{Summary and outlook}

In this paper we discussed the dynamics of a particle whose position and momentum are continuously measured. The detectors, regularly distributed in phase-space, act as a permanent perturbation on the particle. Their effect on the measured system is to collapse the particle wavefunction $\psi(t)$ to a coherent state (the detector "eigenstate") centered at the phase-space point $\alpha = (x_m,k_n)$.  The probability per unit time of a detector click is proportional to the overlap (squared) $|\langle \alpha|\psi(t)\rangle|^2$ of the particle wavefunction and the coherent eigenstate of the detector, in direct analogy to Born's rule.  Likewise, the post-measurement state of the particle is fully determined by the detector-eigenstate. \\

After introducing the model, we  presented a number of (semi-)analytical results explaining several features of the dynamics, coming from different choices of the parameters $\gamma$ and $D$. In particular, we identify the occurrence of a Zeno effect in the limit of a sparse spatial grid of meters and "frequent clicking", $\gamma > v/D$. In this case, the particle's motion (found initially at one of the detectors) is delayed as compared to its free evolution. We showed that the delay time can be explained in terms of renewal theory. Our estimation of the inter-arrival time matches numerical results. Moreover, we showed that if the "measuring intensity" is low, $\gamma < v/D$, the average position of the particle is initially a steplike function of time, in contrast to a dense detector grid, where classical trajectories emerge independent of $\gamma$. \\

Finally we compared our method to an alternative model of continouos measurement, in which the post-measurement state is obtained by applying a Gaussian filter centered around a given position on the wavefunction. For better comparison we assume that these filters are localized at the same discrete points in space as our projective meters. We show that both approaches provide similar results when it comes to the mean position of the particle, however they differ with respect to how the dispersion of position and momentum scale with time.\\

In this study, we assumed that the detectors project on Gaussian states with equivalent width in position and momentum space. One could instead study detectors characterized by squeezed coherent states  $\langle x|\alpha_{m,n}\rangle \propto e^{(x-x_m)^2/(2\sigma^2) +ik_n x}$ -- where $\sigma \neq 1/2$ determines the level of squeezing. By varying $\sigma$ one can smoothly change the relative precision of position and momentum measurement. Likewise, varying the relative spacing in position and momentum of the detectors is another way of further exploring the model. As mentioned at the end of last section, introducing a $\gamma$ parameter which varies between the detectors could also lead to new dynamical features. However, probably the most rewarding extension of this work is to explore an equivalent model in a many-body setting.\\

Commenting on what sets apart the model presented here from the various other continuous measurement schemes studied so far -- and summarized in the introduction of this article -- we would like to point out that we consider a simultaneous measurement of non-commuting observables, that is position and momentum. The result of this measurement operation is the projection on a pre-defined state, in the spirit of Born's rule. This simplicity introduces a potential advantage when it comes to theoretical study of the model: a dynamics which consists of returning every so often to the same Gaussian state (eventually translated in position and/or momentum) may be understood in the context of extended renewal theory. By this we mean that the evolution of the system is characterized completely by functions $\lambda_u (t)$ which denote the overlap squared of a Gaussian detector with a neighbor at a distance $u=(\Delta x, \Delta p)$ in phase space as a function of time. The computational advantage coming from this fact was not explored fully in the present work, since the systems under study were simple enough to be treated within MCWF. Nevertheless we  believe that further investigation in this direction could lead to future valuable insights.\\

\begin{acknowledgments}
The authors acknowledge valuable  discussions and comments of Piotr Szankowski. This work was supported by the Polish National Science Centre  through the
project MAQS under QuantERA, which has received funding from the European
Union’s Horizon 2020 research and innovation program under Grant Agreement
No 731473, Project No 2019/32/Z/ST2/00016.
\end{acknowledgments}

\end{document}